\documentclass[twocolumn, superscriptaddress, secnumarabic, amssymb, nobibnotes, aps, prd, preprintnumbers, nofootinbib]{revtex4-1}

\input{header}

\begin{document}

\preprint{UCI-2025-07}

\title{{\Large Anti-Electron Neutrinos at High-Energy Neutrino Experiments: Identification Strategies and Physics Potential}}

\author{Felix Kling \orcid{0000-0002-3100-6144}}
\affiliation{Department of Physics and Astronomy, University of California, Irvine, CA 92697 USA}
\affiliation{Deutsches Elektronen-Synchrotron DESY, Notkestr.~85, 22607 Hamburg, Germany}

\author{Toni M\"akel\"a \orcid{0000-0002-1723-4028}}
\affiliation{Department of Physics and Astronomy, University of California, Irvine, CA 92697 USA}

\author{Josh McFayden \orcid{0000-0001-9273-2564}}
\affiliation{Department of Physics \& Astronomy, University of Sussex, Sussex House, Falmer, Brighton, BN1 9RH, United Kingdom}

\begin{abstract}
Most existing and proposed high energy neutrino experiments have excellent muon charge identification capabilities, enabling the distinction of $\nu_\mu$ and $\bar \nu_\mu$ charged current interactions. In contrast, distinguishing electrons and positrons from $\nu_e$ and $\bar \nu_e$ interactions is typically impossible, as they quickly interact within the characteristically dense detector material and fail to reach the spectrometer. In this letter, we propose a compact and cost-effective plastic target, placed right before the spectrometer, to maximize the rate of electrons and positrons that reach the spectrometer before interacting. We demonstrate that, when installed at the FASER experiment, the Forward Physics Facility, or SHiP, this setup could enable the first separate measurement of $\nu_e$ and $\bar\nu_e$ cross sections at high energy. Additionally, this setup opens new opportunities to study forward particle production at collider neutrino experiments, such as constraining forward $\Lambda$ hyperon production, and, by reducing flux uncertainties, significantly improve limits on non-standard neutrino interactions in neutral currents.
\end{abstract}

\maketitle 

\noindent \textbf{Introduction -} 
For decades, high energy neutrino interactions have served as a sensitive probe of fundamental physics, for example, to measure neutrino cross sections, constrain proton structure, and search for signs of new physics. In the past, such measurements were performed by accelerator-based neutrino experiments, such as CHARM~\cite{CHARM:1980ppk} or  NuTeV~\cite{Bolton:1990si}, which have probed neutrinos with energies up to several hundred GeV. This program will be continued by the planned SHiP experiment~\cite{Ahdida:2023okr}. 

More recently, new experiments began probing neutrinos with up to multi-TeV energies produced at the Large Hadron Collider (LHC)~\cite{Ariga:2025qup}. In spring 2023, the FASER Collaboration~\cite{FASER:2023zcr} reported the first observation of collider neutrino interactions, which was confirmed shortly after by the SND@LHC Collaboration~\cite{SNDLHC:2023pun}. Since then, FASER has performed the first measurements of neutrino cross sections and fluxes~\cite{FASER:2024hoe, FASER:2024ref}, marking the beginning of a new field of collider neutrino physics. Both experiments plan to continue operations during LHC Run~4 with upgraded detectors~\cite{FASER:2025myb, Abbaneo:2926288}. Additional collider neutrino detectors with significantly increased target mass have also been proposed in the context of the Forward Physics Facility~\cite{Anchordoqui:2021ghd, Feng:2022inv, Adhikary:2024nlv, FPF:2025bor} as well as on the surface exit points of the LHC neutrino beam~\cite{Ariga:2025gtj, Kamp:2025phs}. 

With several new and upgraded high-energy neutrino experiments set to begin operations in the 2030s, it is crucial to explore ways to maximize their physics potential. In this letter, we propose a small plastic target to be placed just before the spectrometer of the planned detectors, enabling the distinction of electron neutrino and anti-neutrino interactions. As we will demonstrate below, this addition will allow separate measurements of their cross-sections at high energies for the first time, provide novel constraints in forward particle production, and enhance the sensitivity of searches for neutrino non-standard interactions (NSI).  \smallskip

\noindent \textbf{Detector Concept -} 
The existing FASER experiment~\cite{FASER:2018ceo, FASER:2018bac, FASER:2019dxq, FASER:2020gpr, FASER:2022hcn} as well as the proposed FASER$\nu$2 and FLArE detectors at the FPF, and the neutrino detector at SHiP, are expected to see a large number of electron, muon and tau neutrino charged current (CC) interactions. All these detectors have excellent charge identification capabilities for the muons from $\nu_\mu$ and $\bar \nu_\mu$ CC interactions, passing through an empty magnetized spectrometer located behind the target volume. However, the distinction of electrons and positrons from $\nu_e$ and $\bar\nu_e$ CC interactions is nearly impossible as the $e^\pm$ interact very quickly within the dense target material, thereby evading charge identification in the spectrometer. 

The exception are electron neutrinos interacting in the last radiation length $X_0$ of the target before the spectrometer - a feature that we aim to exploit by optimizing the corresponding target. The rate of such electron neutrino interactions with escaping electrons is proportional to $\rho X_0 [1-\exp(-L/X_0)]$ with the target's length $L$, and density $\rho$. The product $\rho X_0$ is maximized for materials with low atomic numbers $Z$. Restricting ourselves to solid and non-hazardous materials, the most suitable materials are graphite and plastic. We note that a plastic target has the additional advantage that it can be used as an active scintillator, providing additional information on the vertex position and deposited energy. 

\begin{figure}[tb]
\includegraphics[width=\columnwidth]{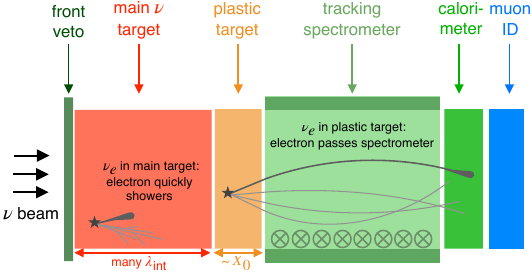}
\caption{Schematic of the setup, illustrating the position of the proposed plastic target and the path of the electrons and positrons through the experiment.}
\label{fig:schema}
\end{figure} 

\begin{table}[tb]
\setlength{\tabcolsep}{6pt}
\centering
\begin{tabular}{l  c c   c c c}
\hline\hline
 & \multicolumn{2}{c}{main detector} 
 & \multicolumn{3}{c}{plastic target}\\
 & mass & $\nu_e + \bar\nu_e$ 
 & mass & $\nu_e$ & $\bar\nu_e$\\
\hline
 FASER & 1.2~ton & 4.9k & 6.5~kg & 21 & 11 \\
 FPF   & 20~ton & 210k & 600~kg & 1120 & 525 \\
 SHiP  & 3~ton & 3.3M & 33~kg & 30k & 6.6k \\
\hline\hline                     
\end{tabular}
\caption{Target mass and expected number of electron neutrino interactions in FASER with $680~\ifb$, the FPF with $3~\iab$ and SHiP with $6 \cdot 10^{20}$ protons on target.}
\label{tab:rates}
\end{table}

The term $1-\exp(-L/X_0)$ is close to be maximized for $L \sim X_0$, corresponding to target length of $20$ and $40~\cm$ for graphite and plastic, respectively. This coincides with the space available at the current FASER location, without excessive compromise on the size of the main neutrino target. In the following, we therefore consider the addition of a 20~cm thick plastic target in front of the existing FASER spectrometer. We also consider a 20 cm thick target for FASER2 and SHiP. However, spatial constraints are significantly relaxed at these locations, allowing for the possibility of deeper detectors to increase event statistics. A schematic experimental setup is illustrated in \cref{fig:schema} using the example of the FASER experiment. The information of the tracker and recently installed high-granularity pre-shower detector~\cite{Boyd:2803084} and muon system then allow to reliably identify the electron track and measure its charge and momentum. 

The considered target masses of the experiment's main detector and the proposed plastic targets, along with the expected number of electron neutrino interactions for all detectors, are summarized in \cref{tab:rates}. This estimate is based on the flux predictions presented in Ref.~\cite{Kling:2021gos, Buonocore:2023kna, FASER:2024ykc} for the LHC and in Ref.~\cite{Albanese:2878604, Ahdida:2023okr} for SHiP. As we will see below, selection cuts are not expected to significantly reduce the observable event rate. \medskip

\noindent \textbf{Signal Identification -} 
To illustrate the feasibility of a separate $\nu_e$ and $\bar\nu_e$ measurements with the proposed detector design, we perform a simulation to study signal identification and background rejection. For concreteness, we consider a 20~cm thick plastic target at the FPF, but note that the same arguments also apply for FASER and SHiP. 

In order to consider the efficiency of identifying the $\nu_e$CC signal from the background $\nu_\mu$CC and neutral current (NC) events at FASER2, a sample containing $\nu_e$CC and $\overline{\nu}_e$CC signal events and $\nu_\mu$CC and NC background events is generated using \textsc{Pythia8}~\cite{Bierlich:2022pfr}. To simulate the electromagnetic (EM) showers in the plastic detector, the photons and $e^\pm$ in the generated sample are subjected to a showering algorithm accounting for $\gamma \to e^+e^-$ pair production following Ref.~\cite{ParticleDataGroup:2024cfk}, while the generation of additional photons in $e^\pm \to e^\pm + \gamma$ bremsstrahlung is implemented following Ref.~\cite{Klein:1998du, PhysRevD.25.1291}, accounting for the Landau-Migdal-Pomeranchuk effect~\cite{Landau:1953ivy, Landau:1953gr, PhysRev.103.1811} and avoiding the $x=0$ singularity of the Bremsstrahlung cross section in Ref.~\cite{ParticleDataGroup:2024cfk}. The algorithm is recursively applied on the additional photons and $e^\pm$, until the energy of the produced particles becomes negligible or the shower reaches the end of the detector volume. The implementation of the procession of the EM shower through the detector is inspired by Ref.~\cite{Salvat2019PENELOPE2018}\footnote{Ref.~\cite{Salvat2019PENELOPE2018} uses a different integration variable $W$,  covering all energies from a low-$E$ cutoff to the energy of the radiated photon. Here, $dx$ is used instead of $dW$, but any factors introduced by the change of variables cancel, while the integration limits become from 0 to 1, and the normalization is taken to be the radiation length.}. The inverse of the mean free path (MFP) for a particle of energy $E$, interacting via a process with a differential cross section $d\sigma/dx$, reads
\begin{equation}
\lambda^{-1}(E)
=
\frac{1}{X_0}
\int_{0}^{1}
\frac{d\sigma(E)}{dx}
dx.
\end{equation}
For pair production, this yields $\lambda \approx 9/7 X_0$, as expected. For Bremsstrahlung however, the MFP is significantly below $X_0$. This determines the length scale required in sampling a longitudinal step size from the exponential distribution. Transverse smearing is added by sampling a squared transverse momentum for the produced particles according to $p_T^2 = b((1 + p_{T{\rm ,max}}^2/b)^u - 1)$, which essentially inverts the cumulative distribution function of $1/(p_T^2 + b)$. Here, $u \in {\rm Uniform}[0,1]$, $p_{T{\rm ,max}} = x E \theta_{\rm max}$ and $b=x^2 m_e^2$ ($b=m_e^2$) for Bremmsstrahlung (pair production), with $\theta_{\rm max}=0.1$ based on the small angle approximation being good for angles below 0.1 rad.\medskip

We assume that the detector has good capabilities for measuring the EM energy $E_{\rm EM}$ arising from photons and $e^\pm$, the energy of all visible particles $E_{\rm Tot}$, as well as the energies of the most energetic muons and electrons, $E_\mu$ and $E_{e^\pm}$, respectively, as well as their charge. 
To contribute to these observables, charged particles are required to have energies above 20~GeV in order to pass through the whole spectrometer. 

Using the above observables and the simple physics-motivated cuts listed in \cref{tab:cutflow} allows a good separation of the signal and SM backgrounds already in a cut and count analysis. Over 99\% of all $\nu_\mu$CC backgrounds are removed with minimal effect to the signal by demanding that there are no muons with energies above 20~GeV. After this, it remains to distinguish $\nu_e$CC events from NC events producing electrons. Demanding that the $E_{\rm EM}$ in the event is above 40\% of the energy of all visible particles in the event removes 40\% of NC events. Requiring the highest electron or positron energies to be above 25\% of the EM energy removes another 50\% of the remaining NC event, particularly those occurring in the back of the detector. The majority of the remaining backgrounds is removed by restricting to events with an electron or positron of at least a 100~GeV, which however also annuls all signal events in the $\mathcal{O}$(10~GeV) range.

\begin{table}[t]
\setlength{\tabcolsep}{6pt}
\centering
\begin{tabular}{c c c c c}
    \hline\hline
    Cut & $\nu_e$CC & $\overline{\nu}_e$CC & $\nu_\mu$CC & NC \\
    \hline
    No cuts
    & 1120 & 525 & 8298 & 243 \\
    $E_\mu < 20$~GeV
    & 1111 & 520 & 34 & 235 \\
    $E_{\rm EM}/E_{\rm Tot} > 40$\%
    &  984 & 490 & 22 & 144 \\
    $E_{e^\pm}/E_{\rm EM} > 25$\%
    & 856 & 434 & 15 & 73 \\
    $E_{e^\pm}>100$~GeV 
    & 612 & 332 & 1 & 19 \\
    \hline
    BDT & 1075 & 515 & 13 & 112\\
    \hline\hline
\end{tabular}
\caption{Event statistics after simple physics-motivated cuts that could be used for a cut and count analysis to distinguish the signal $\nu_e$CC events from the background $\nu_\mu$CC and NC events. The numbers of events of each category classified as signal events in the BDT analysis are shown for comparison. The BDT preserves the signal well across all energies, at the expense of retaining more NC events than the final cuts in the cut and count analysis, which also remove signal events. See text for details.}
\label{tab:cutflow}
\end{table}

\begin{figure}[tbh]
\includegraphics[width=0.47\textwidth]{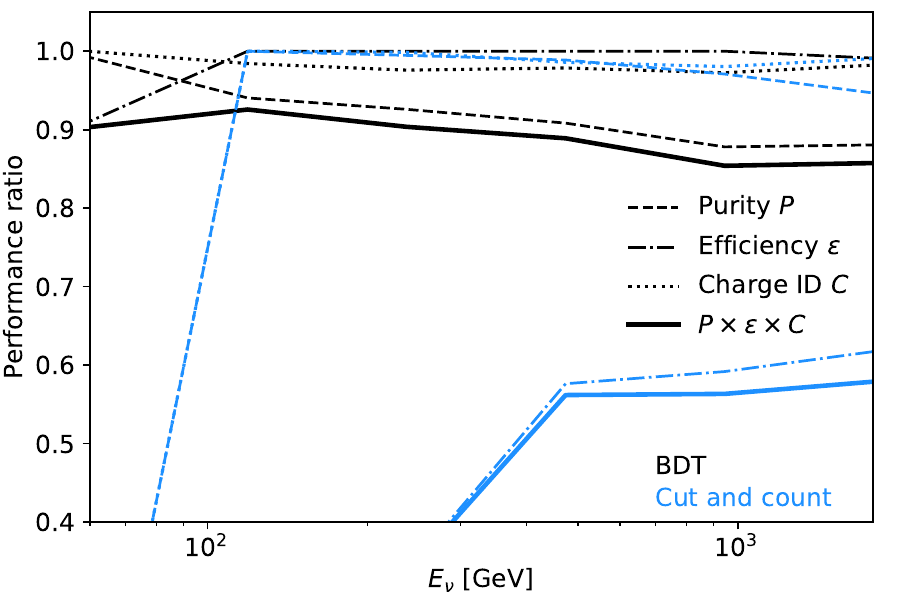}
\caption{The energy dependence of the detector performance metrics for a 20 cm detector. The result of the BDT analysis (black) is contrasted with the result of a simple cut and count analysis (blue).}
\label{fig:performance_ratios_fix_depth}
\end{figure}

\begin{figure}[tbh]
\includegraphics[width=0.47\textwidth]{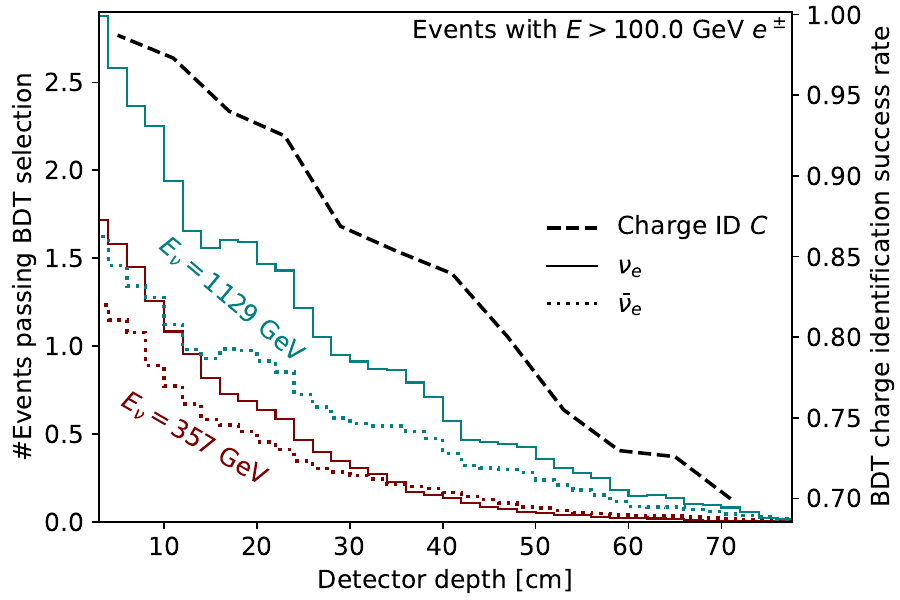}
\caption{The events passing the BDT selection criteria and containing an electron of $E_{e} >100~\gev$, originating at a given depth in the detector, for neutrino energies of 357 GeV (brown), and 1129 GeV (teal). The solid (dashed) lines correspond to $\nu_e$CC ($\overline{\nu}_e$CC) events, and the numbers of events originating at a given depth are given on the left vertical axis. Detectors deeper than 50 cm add no significant contributions to the number of signal events with a high-energy electron. The black dashed line indicates the BDT charge identification success rate, with values given on the right vertical axis.}
\label{fig:depth_optimization}
\end{figure}

The signal identification performance is assessed by considering the sample purity $P$, defined as the ratio of $\nu_e+\overline{\nu}_e$ signal events to all signal and background events; the charge identification success rate $C$, defined as the ratio of neutrino and antineutrino-induced events labeled correctly as $\nu_e$CC or $\overline{\nu}_e$CC events vs all signal events surviving the cuts; and efficiency $\epsilon$, defined as the fraction of signal events after and before the cuts. In the cut and count analysis, events are classified as originating from $\nu_e$ or $\overline{\nu}_e$ depending on the sign of the most energetic $e^\pm$, and a sample purity above 90\% and near-perfect charge identification can be achieved, retaining more than 50\% of the signal events with $E_\nu$ above a few hundred GeV, as illustrated in \cref{fig:performance_ratios_fix_depth}.

The signal acceptance and background rejection can be improved further by an optimized multivariate analysis considering additional correlations between the observables. To this end, the observables listed above are passed as features to the boosted decision tree implemented in the \textsc{scikit-learn} package~\cite{scikit-learn}, tasked with classifying events in the generated sample into $\nu_e$CC, $\overline{\nu}_e$CC, $\nu_\mu$CC and NC categories. 
Besides the observables utilized in the cut and count analysis, the BDT is passed the following additional features: the neutrino interaction vertex position achievable via a longitudinally segmented plastic target, the energy of all charged tracks $E_{\rm Ch}$, the total charge of $e^\pm$ with energies above 100 GeV, the leading $e^\pm$ angle with respect to the line of sight, as well as the standard deviation and maximum of the angles of all EM tracks from the leading $e^\pm$, characterizing the spread of the EM shower. This results in an excellent performance, with the product of the sample purity, charge identification rate, and efficiency, $P\times C \times \epsilon$ around 90\% in the $E_\nu\sim\mathcal{O}$(100~GeV) range, as shown in \cref{fig:performance_ratios_fix_depth}.

If additional space is available, a deeper plastic target can be considered for increasing event statistics. In the following, we consider the performance metrics in terms of the longitudinal vertex location to estimate an optimal target depth. \cref{fig:depth_optimization} illustrates the number of events passing the BDT selection cuts and the associated BDT charge identification success rate as a function of the vertex $z$ position. While increasing the detector depth will in general increase statistics, the performance metrics decrease with increasing $z$ such that detector depths exceeding about 50 cm bring no significant additions to the signal. Therefore, if space and longitudinal segmentation are available, a detector depth corresponding to $X_0$ could be considered for increasing statistics, but deeper targets will provide diminishing returns.

Finally, neutrino energy resolution is estimated by performing a regression BDT analysis, using the same observables as features as the BDT analysis discussed above. This time, the BDT is trained to reconstruct the incoming neutrino energies in events classified as $\nu_e$CC or $\overline{\nu}_e$CC. As a result, a linear dependence between the reconstructed and true $E_\nu$ is observed, with a relative standard error of $\delta E_\nu/E_\nu < 30$\% for the reconstruction.\medskip

\begin{figure}[t]
\includegraphics[width=\columnwidth]{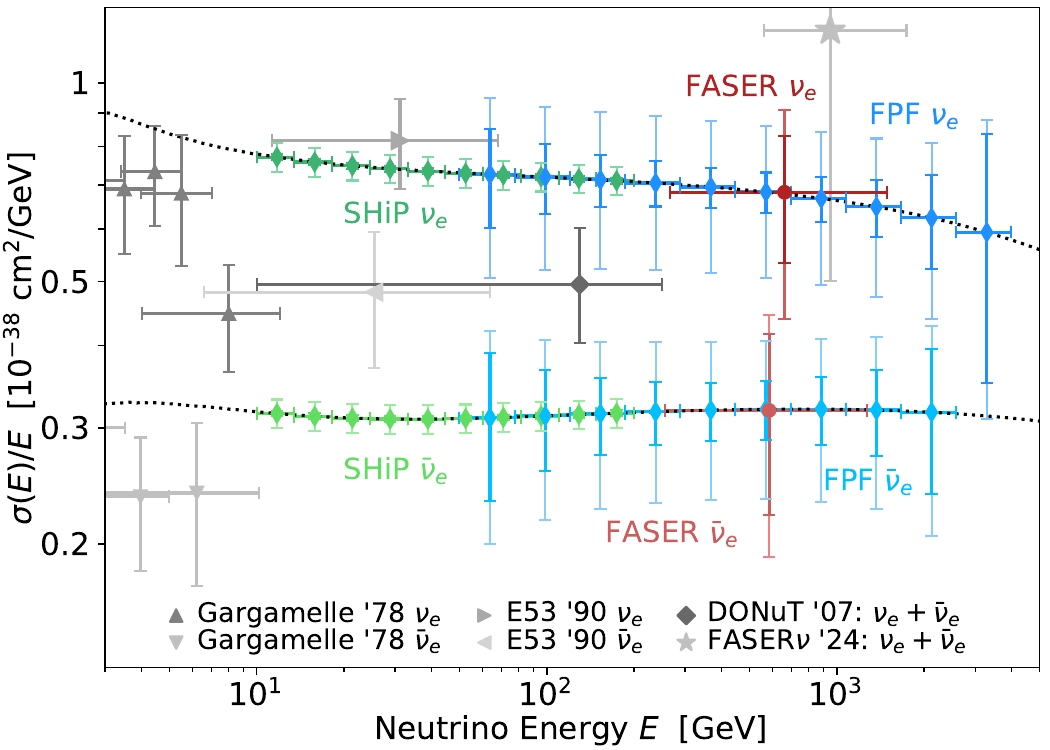}
\caption{Predicted precision of the neutrino-nucleon cross sections for $\nu_e$ and $\bar\nu_e$, normalized to the neutrino energy, using a plastic target at FASER, the FPF and SHiP. We also show previous measurements performed at Gargamelle~\cite{Gargamelle:1977pmg}, E53~\cite{Baltay:1988au}, DONuT~\cite{DONuT:2007bsg} and FASER$\nu$~\cite{FASER:2024ref}. The dotted line shows the predicted cross section obtained using the Bodek-Yang model~\cite{Bodek:2002vp} as implemented in GENIE~\cite{Andreopoulos:2009rq, GENIE:2021npt}. See text for details.}
\label{fig:xs}
\end{figure} 

\noindent \textbf{Neutrino Cross Sections -} 
Although the number of total events in this plastic target is far below that of the main neutrino targets, it offers several novel measurement opportunities. This includes the separate measurement of the $\nu_e$ and $\overline{\nu}_e$ interaction cross section as well as their fluxes. We note that this detector alone will only constrain a combination of neutrino flux and interaction cross section. However, this degeneracy can be broken using external input: prior knowledge on one quantity allows us to interpret the data in terms of the other. 

Our first example is the separate measurement of the $\nu_e$ and $\overline{\nu}_e$ interaction cross section. While separate cross section measurements have been performed for muon neutrinos and anti-neutrinos by a large number of experiments, mainly combined measurements are available for high-energy electron neutrinos and anti-neutrinos as performed by DONuT~\cite{DONuT:2007bsg} and FASER$\nu$~\cite{FASER:2024ref}. Separate measurements at low energies have been performed using the bubble chamber experiments Gargamelle~\cite{Gargamelle:1977pmg} and E53~\cite{Baltay:1988au}, partially using separate neutrino and anti-neutrino dominated beams. 

In Fig.~\ref{fig:xs}, we present the projected sensitivity to separately measure the $\nu_e$ and $\bar{\nu}_e$ cross sections using a plastic target at FASER, the FPF, and SHiP. The inner (thicker) error bars indicate the expected statistical precision, corresponding to uncertainties of approximately 25\%, 10\%, and 1\% per bin at FASER, the FPF, and SHiP, respectively. The outer (thinner) error bars include an approximate estimate of the flux uncertainties. At SHiP, the neutrino flux is predominantly produced in the decays of charmed hadrons. Their production rate has been constrained by the NA27 experiment with a precision of about 20\%~\cite{LEBC-EHS:1988oic}, and the ongoing NA65 experiment is expected to further reduce this uncertainty to a few percent~\cite{DsTau:2019wjb}. For illustration, we therefore assume a 5\% flux uncertainty for SHiP in Fig.~\ref{fig:xs}. At the LHC, the neutrino flux uncertainty is primarily driven by the modeling of forward hadron production in hadronic interaction models at collider energies. Current state-of-the-art frameworks such as EPOS, QGSJET, and SIBYLL still exhibit sizable differences at the level of about 25\%~\cite{FASER:2024ykc}. A comparable uncertainty was also obtained from tuning variations in Pythia in Ref.~\cite{Fieg:2023kld}. A concerted community effort to improve hadronic interaction modeling is ongoing and is expected to further reduce these uncertainties by the time such measurements are performed. \medskip

\noindent \textbf{Forward Hyperon Production -} The LHC neutrino beam is mainly generated via the decays of forward charged pions, kaons, charm hadrons, and hyperons. The production of these particles in the forward region at LHC energies has not been measured before. Thus, collider neutrino flux measurements offer unique constraints on forward particle production, complementing measurements of $\pi^0$, $\eta$ and neutron production by LHCf~\cite{LHCf:2017fnw, LHCf:2018gbv, Piparo:2023yam}. These data are crucial for improving and validating hadronic interaction models, which are particularly important in astroparticle physics to model particle production in extreme astrophysical environments and describe cosmic ray interactions in the Earth's atmosphere.

At the highest rapidities, electron neutrinos and anti-neutrinos primarily originate from kaons and hyperons at energies $E_\nu<\tev$, as well as charm hadrons at higher energies. As discussed in Ref.~\cite{Kling:2021gos}, the fluxes of $\nu_e$ and $\bar\nu_e$ are not equal, which is a consequence of the net charge and baryon number of the proton-proton collision system. This leads, for example, to a strongly enhanced production of forward $\Lambda$ baryons compared to a negligible rate of $\bar\Lambda$ anti-baryons, and hence induces an enhancement of $\bar\nu_e$ neutrino flux via the decay $\Lambda \to p e^- \bar \nu_e$. 

This feature enables a novel measurement of forward $\Lambda$ production. First, note that the production of charged kaons, and the resulting $\nu_e$ and $\bar\nu_e$ flux from their decays, can be tightly constrained using $\nu_\mu$ and $\bar\nu_\mu$ flux measurements at the main target of the collider neutrino experiment. The remaining, unconstrained component of the flux then primarily originates from $K^0$ and $\Lambda$ decays. Since $K^0$ decays produce $\nu_e$ and $\bar\nu_e$ symmetrically, while $\Lambda$ decays yield only $\bar\nu_e$, any $\bar\nu_e - \nu_e$ flux asymmetry observed using the plastic target can be attributed to $\Lambda$ decays.

\begin{figure}[tbh]
\includegraphics[width=\columnwidth]{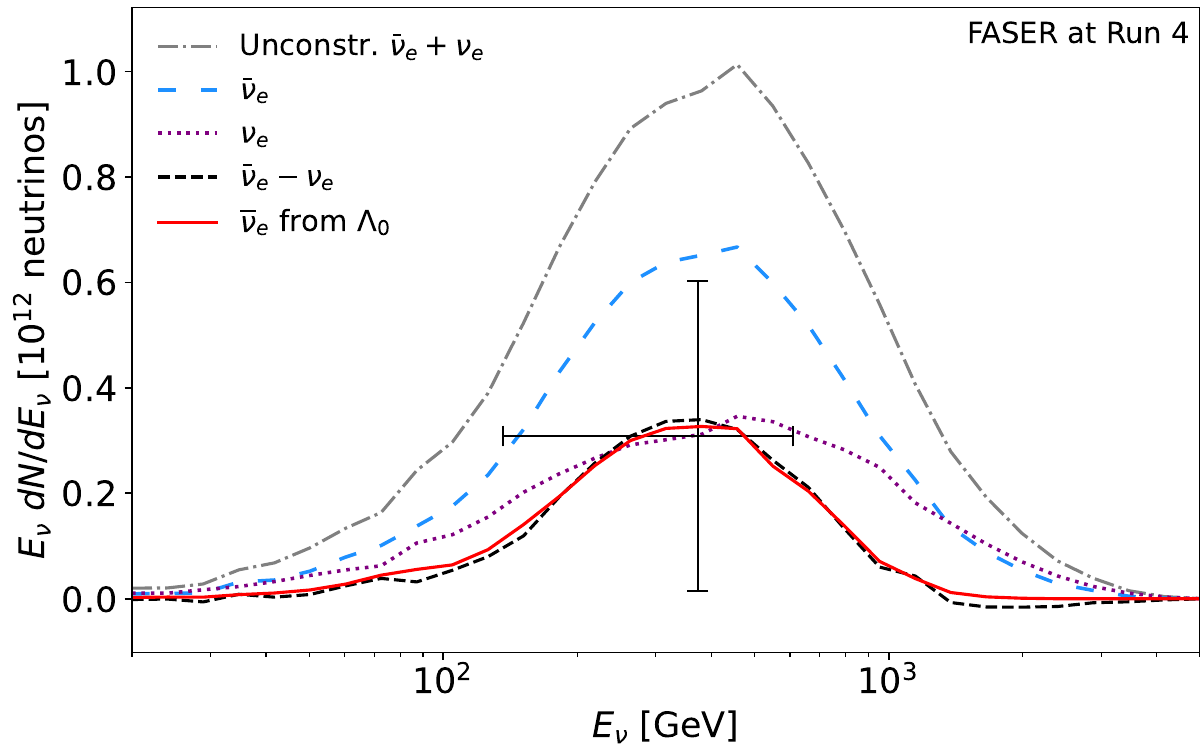}\\
\includegraphics[width=\columnwidth]{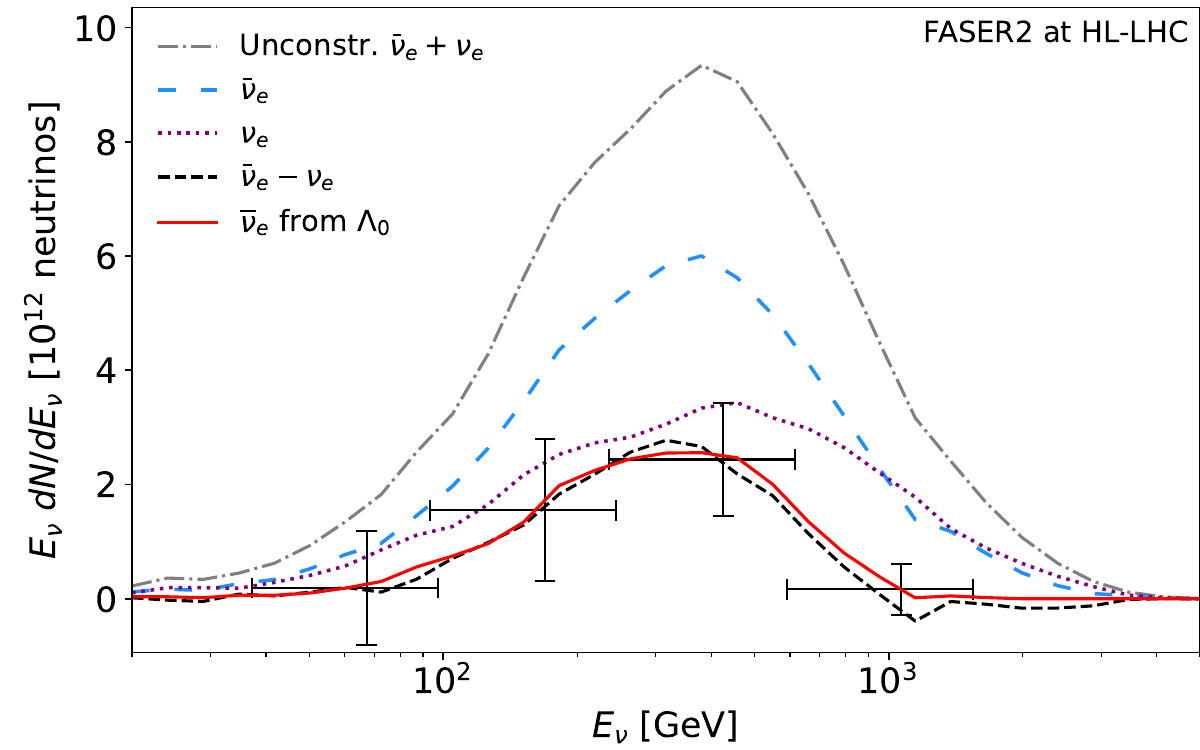}
\caption{Flux of $\nu_e$, $\bar\nu_e$, their sum, and their difference, originating from the unconstrained $K^0$, $\Lambda$ and charm hadrons passing through the plastic target at FASER and the FPF. The neutrino flux from $\Lambda$ decays is shown for comparison. The errorbars depict the expected precision of its measurement as inferred from the statistical uncertainties in the number of $\nu_e$ and $\bar\nu_e$ CC interaction events in the plastic target.}
\label{fig:Nlambda0}
\end{figure} 

This is illustrated in \cref{fig:Nlambda0}, which shows the $\nu_e$ and $\bar\nu_e$ fluxes, along with their sum and difference, at FASER and the FPF, as estimated using \texttt{SIBYLL~2.3d}~\cite{Riehn:2019jet} and \texttt{POWHEG+Pythia}~\cite{Buonocore:2023kna}. At the FPF, we only consider the rapidity region $\eta>8.7$, since the $\Lambda$ component decreases for lower rapidities. We can see that the $\bar\nu_e-\nu_e$ flux difference (black dashed) is indeed a good estimator of the flux of neutrinos from $\Lambda$ decays (red solid), hence confirming the conceptual argument. The estimated precision at which the neutrino flux from $\Lambda$ decay is shown via the errorbars. Here we assume SM neutrino scattering cross sections which are constrained at the few percent level, with the uncertainties arising predominantly from the parton distribution functions. While a first measurement, statistically limited, is possible at FASER in Run~4, a differential analysis becomes feasible at the FPF. \medskip

\noindent \textbf{Neutrino Non-Standard Interactions -} The ability of $\nu_e$ and $\bar\nu_e$ separation via the plastic target will also contribute to reducing systematic uncertainties in measurements performed with the main targets of collider neutrino detectors. We demonstrate this by considering a measurement of neutrino NSI, which are parameterized by dimension-6 effective field theory operators
\be
\nonumber
\!\!\mathcal{L}
\!=\! -\sqrt{2} G_F
[\overline{\nu}_\alpha\gamma^\mu P_L \nu_\beta]
[\epsilon^{qV}_{\alpha\beta} \overline{q}\gamma_\mu q 
\!+\!\epsilon^{qA}_{\alpha\beta} \overline{q}\gamma_\mu\gamma^5 q].
\ee
Here, $G_F$ is the Fermi constant and $\epsilon^{q,V/A}_{\alpha\beta}$ are Wilson coefficients for operators modifying the (axial) vector coupling between neutrinos and a quark $q$. Alternatively, one can consider the operators affecting the left- and right handed couplings by writing the $\epsilon^{qV/A}_{\alpha\beta}$ in terms of $\epsilon^{qL/R}_{\alpha\beta}$.

The presence of such NSI modify the rate of NC interactions, which can be measured with high statistics in the main targets. Conceptually, one can use the CC interaction rate measurement to constrain the incoming neutrino flux and then search for deviations in the NC interaction rate due to NSI. For illustration, we focus on NSI affecting muon neutrinos, which are most abundant at colliders and whose flux will be well constrained. However, since neutrino flavors are indistinguishable in NC events, uncertainties in the $\nu_e$ and $\bar\nu_e$ fluxes will propagate into the NC rate measurement and weaken the sensitivity to NSI. 

To quantify this effect, we use simulated data to derive projected 95\% confidence level bounds on the NSI Wilson coefficients. We perform a binned $\chi^2$ analysis based on the reconstructed neutrino energy distribution, incorporating both statistical and systematic uncertainties in each energy bin, added in quadrature. As a validation step, we reproduce the results of Ref.~\cite{Ismail:2020yqc}, which assumed only statistical uncertainties and included only muon (anti)neutrinos in the incoming flux. Our analysis generalizes this by accounting for all neutrino flavors contributing to the observed NC interactions and by including flux uncertainties.

The systematic uncertainties for the electron neutrino contribution are estimated for scenarios with and without input from a plastic target. While the total rate of $\nu_e+\bar\nu_e$ CC interactions is well constrained using measurements at the main target, the relative contributions from $\nu_e$ and $\bar\nu_e$ remain unconstrained. To conservatively parameterize the resulting flux uncertainty in the absence of a plastic target, we assume that the unconstrained part of the electron neutrino spectrum could originate entirely from either $K^0$ or $\Lambda$ decays. The resulting variation on the predicted number of NC interactions defines the systematic uncertainty. If a plastic target is integrated into the experiment, the uncertainties of the electron neutrino flux and NC interaction rate can be reduced. We obtain an estimate by constraining the hyperon fraction to be consistent with the projected measurement, as discussed in the previous section. 

\begin{figure}[htp!]
\centering
\includegraphics[width=0.47\textwidth]{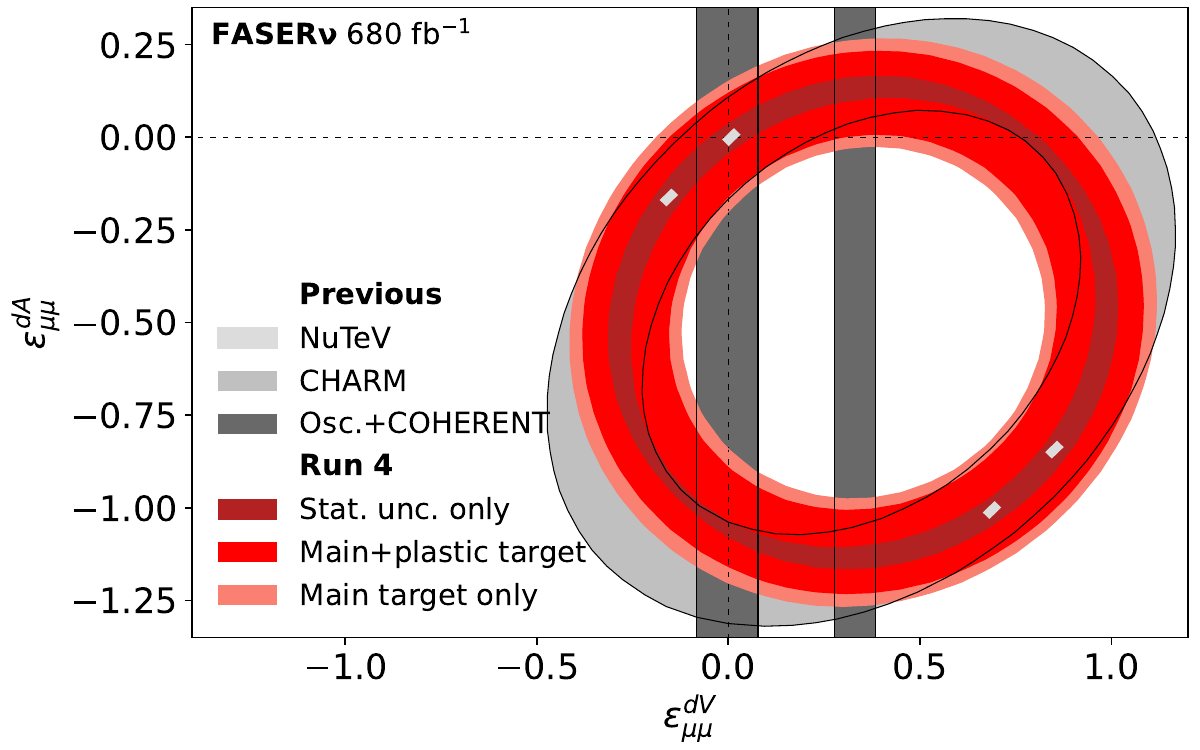}\\
\includegraphics[width=0.47\textwidth]{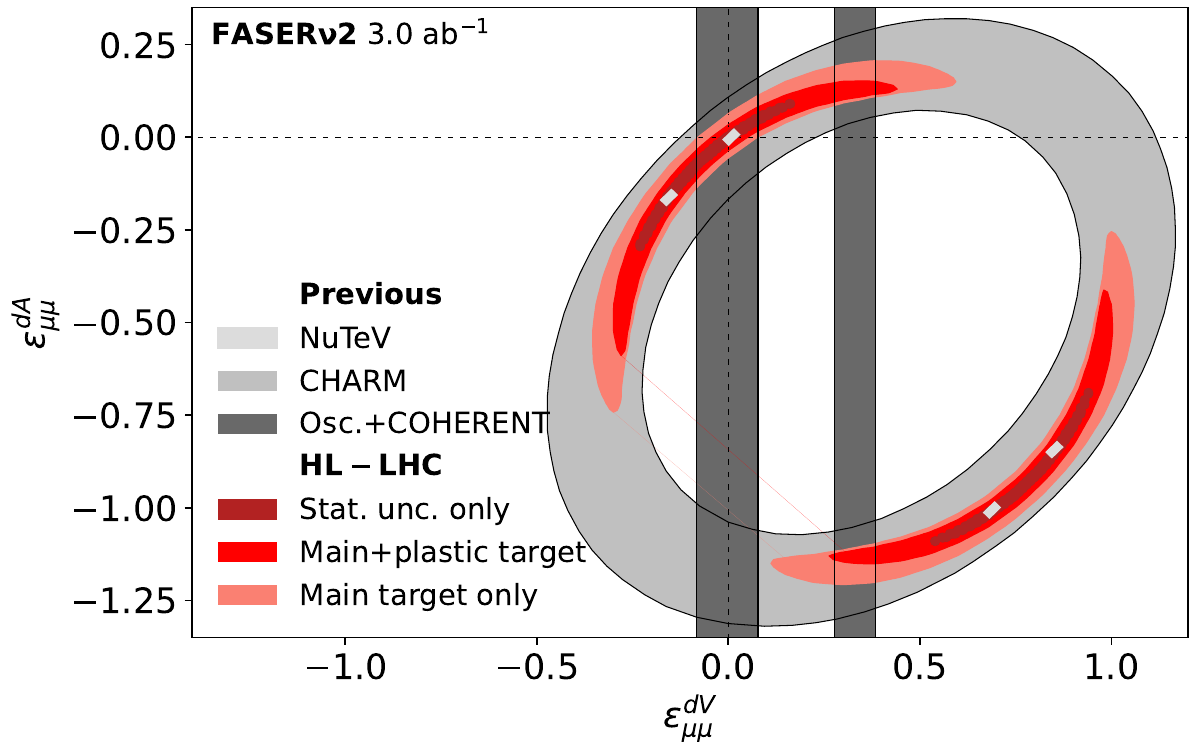}
\caption{The 95\% CL bounds on the Wilson coefficients $\epsilon^{dV}_{\mu\mu}$ and $\epsilon^{dA}_{\mu\mu}$ at FASER (top) and at the FASER2 (bottom) assuming flux uncertainties constrained by the main target (light red) as well as the main and plastic target (red). The sensitivity considering only statistical uncertainties is shown for comparison (dark red). Previous constraints are shown in gray~\cite{CHARM:1987pwr, Coloma:2019mbs, NuTeV:2001whx, Davidson:2003ha}. The intersection of the black dotted lines indicates the SM point.}
\label{fig:NSI_dVdA}
\end{figure}

\begin{figure}[htp!]
\centering
\includegraphics[width=0.48\textwidth]{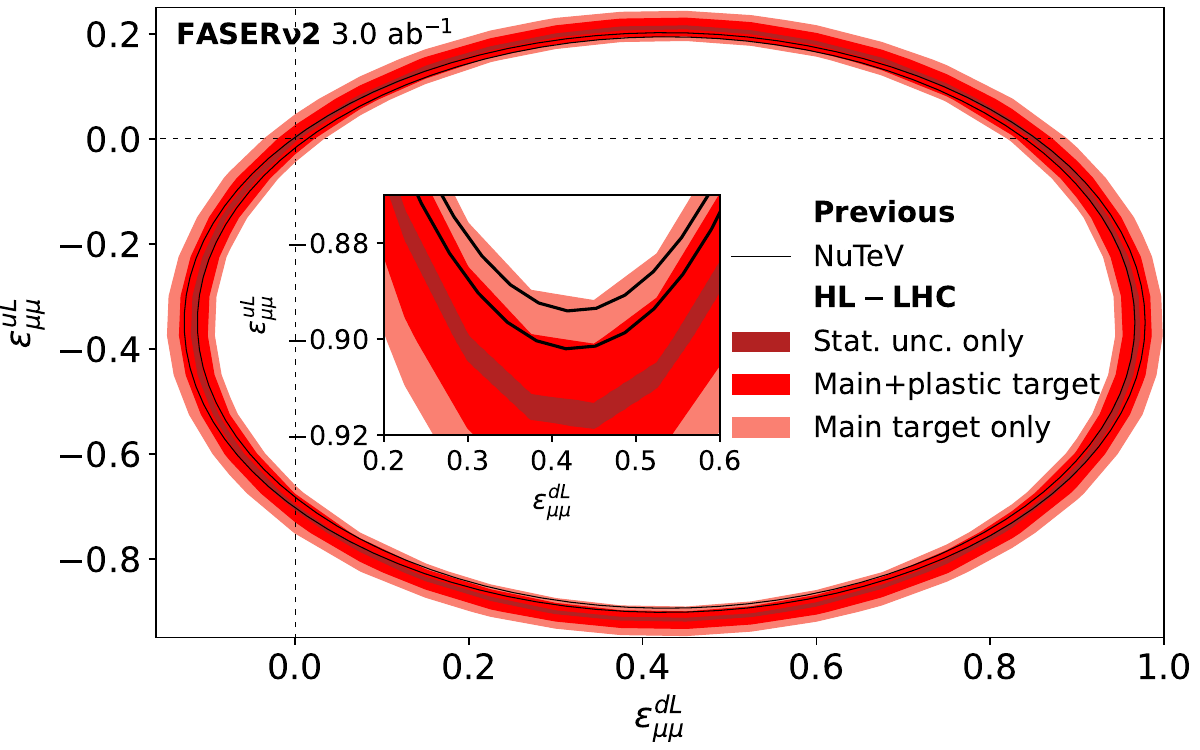}
\caption{The 95\% CL bounds on the Wilson coefficients $\epsilon^{dL}_{\mu\mu}$ and $\epsilon^{uL}_{\mu\mu}$ at FASER2 assuming flux uncertainties constrained by the main target (light red) as well as the main and plastic target (red). The sensitivity considering only statistical uncertainties is shown for comparison (dark red). The area between black ellipses is allowed by previous constraints from NuTeV~\cite{NuTeV:2001whx, Davidson:2003ha}. The insert indicates a region where FASER2 can complement the stringent NuTeV results, if the plastic target is included. The black dotted lines intersect at the SM point.}
\label{fig:NSI_dLuL}
\end{figure}

The projected bounds on two Wilson coefficients, $\epsilon^{dV}_{\mu\mu}$ and $\epsilon^{dA}_{\mu\mu}$ are shown in \cref{fig:NSI_dVdA}, with all other NSI parameters assumed to vanish. Existing constraints were derived from a global fit to neutrino oscillation and COHERENT data ~\cite{Coloma:2019mbs}, and from neutrino scattering measurements at the CHARM experiment~\cite{CHARM:1987pwr} and the NuTeV experiment~\cite{NuTeV:2001whx, Davidson:2003ha}. While the neutrino oscillation and COHERENT data are only sensitive to the vector coupling, but not axial counterparts, high-energy neutrino scattering experiments can probe NSI regardless of the underlying spin structure. The constraints from NuTeV are particularly stringent due to the ability to perform separate $\nu$ and $\bar\nu$ measurements.

Even without the plastic target, measurements at the main targets already provide complementary constraints to the constraints from CHARM as well as neutrino oscillations and COHERENT. Incorporating the plastic target significantly improves sensitivity by reducing flux-related systematics. We note that the measurement remains limited by flux systematics, as demonstrated by the comparison to the contour assuming only statistical uncertainties. The same conclusions hold for measurements at the FPF. 

The constraints from the NuTeV measurement~\cite{NuTeV:2001whx} were however originally presented in terms of the left- and right handed coefficients and varying both $u$ and $d$ quark coefficients simultaneously in Ref.~\cite{Davidson:2003ha}. The projected bounds for $\epsilon^{dL}_{\mu\mu}$ and $\epsilon^{uL}_{\mu\mu}$ at the FPF are shown in \cref{fig:NSI_dLuL} and compared to the constraints from NuTeV. In this parameter space, there is a region where the FPF, with the proposed plastic scintillator included, may complement the NuTeV constraints due to the mildly different shapes of the elliptical constraint bands due to differences in the numbers of neutrons and protons in the target nuclei of the two experiments. \medskip

\noindent \textbf{Conclusions -} 
The recently initiated forward neutrino program at the LHC has begun to provide the first observations and measurements, and the future is promising: both the FASER and SND@LHC Collaborations will continue operations during the LHC Run 4, the SHiP experiment has been approved, and the proposed FPF is projected to yield an abundance of precision results for physics both within and beyond the SM. 
In order to make the best use of this unique direction of LHC physics, it is important to consider how to develop the detection capabilities of these experiments even further. 

In the present work, a small plastic target placed just before the spectrometer of the planned detectors is proposed, allowing the distinction of $\nu_e$ and $\overline{\nu}_e$ interactions. 
It would yield the first separate measurements of their cross-sections at high energies, and help reduce the electron neutrino flux uncertainty. Furthermore, as the contribution of hyperons to the forward hadron flux can be understood through a measurement of the difference of the $\overline{\nu}_e$ and $\nu_e$ spectra, this allows constraining forward strangeness in the baryon sector, providing results that are useful also for the astrophysics community.

When installed at FASER or the at proposed FPF, the considered detector is demonstrated to reduce the uncertainties in the electron neutrino flux and the spectrum of NC interactions, leading to enhanced sensitivity for neutrino NSI in NC interactions. Hence, despite its simple and cost-effective design, the proposed detector can increase the physics potential of existing and proposed forward neutrino experiments at the LHC, SHiP, and potential future detectors at the LHC and beyond~\cite{MammenAbraham:2024gun, InternationalMuonCollider:2025sys}. \medskip

\noindent \textbf{Acknowledgements -} 
We thank Jamie Boyd, Sebastian Trojanowski and Max Fieg for useful discussions.
We are grateful to the authors and maintainers of many open-source software packages, including \textsc{Rivet}~\cite{Buckley:2010ar, Bierlich:2019rhm}, \textsc{scikit-hep}~\cite{Rodrigues:2020syo} and \textsc{scikit-learn}~\cite{scikit-learn}.
The work of F.~K. is supported in part by Heising-Simons Foundation Grant 2020-1840, by U.S. National Science Foundation Grant PHY-2210283 and PHY-2514888, and by the Deutsche Forschungsgemeinschaft under Germany's Excellence Strategy -- EXC 2121 Quantum Universe -- 390833306.
The work of T.M. is supported in part by U.S.~National Science Foundation Grants PHY-2111427 and PHY-2210283 and Heising-Simons Foundation Grant 2020-1840, and by the National Science Centre, Poland, Research Grant No. 2021/42/E/ST2/00031.
The work of JM was supported by the Royal Society grant URF\textbackslash R1\textbackslash201519 and STFC grant ST/W000512/1.

\bibliography{references}

\end{document}